\pdfoutput=1
\documentclass{article}
\usepackage{spconf,amsmath,amssymb,graphicx}
\usepackage{algorithm}
\usepackage{algpseudocode}
\usepackage[skip=2pt]{caption}
\usepackage{subcaption}
\usepackage[hidelinks]{hyperref} % must be the last package

\title{Spatially Filtered Sparse Bayesian Learning for Direction-of-Arrival Estimation with Leaky-Wave Antennas}

\name{{\bf R.Maydani}$^{1}$ {\bf and Y. Wang}$^{1}$, {\bf J. Sarrazin}$^{2}$, {\bf B. Ma}$^{3}$,}
\address{
$^{1}$Nantes Université, CNRS, IETR, UMR 6164, F-44000 Nantes, France\\
$^{2}$Université Paris-Saclay, CentraleSupélec, CNRS, GeePs, 91192, Gif-sur-Yvette, France\\
$^{3}$ School of Electronics and Information Engineering, South China University of Technology,\\ 510640 Guangzhou, China}

\begin{document}
%\ninept
%
\maketitle
\begin{abstract}
Direction-of-arrival (DoA) estimation with leaky-wave antennas (LWAs) offers a compact and cost-effective alternative to conventional antenna arrays but remains challenging in the presence of coherent sources. To address this issue, we propose a spatially filtered sparse Bayesian learning (SF-SBL) framework. Firstly, the field of view (FoV) is divided into angular sectors according to the frequency beam-scanning property of LWAs, and Bayesian inverse problems are then solved within each sector to improve efficiency and reduce computational cost. Both on-grid SBL and off-grid SBL formulations are developed. Simulation results show that the proposed approach achieves robust and accurate DoA estimation, even with coherent sources.
\end{abstract}


\section{Introduction}
\label{sec:intro}

Direction-of-arrival (DoA) estimation is essential in radar, wireless communications, and sensing, but conventional implementations typically require many radio-frequency (RF) channels, which increases hardware cost and system complexity~\cite{2009-Tuncer,doa}. Frequency beam-scanning leaky-wave antennas (LWAs)~\cite{LWAbook} provide a hardware-efficient alternative by generating frequency-dependent beams without phase shifters or complex feeding networks. In particular, single-beam LWAs can cover a wide field of view (FoV) with a single aperture, making them attractive for compact DoA estimation systems~\cite{Gil-2022,2017-Husain,2019-Poveda}.

Since their introduction in DoA estimation~\cite{2012-Yu}, LWAs have attracted growing attention as cost-effective candidates for high-resolution angle finding. For single-beam LWAs, subspace-based algorithms such as MUSIC can provide accurate estimation for incoherent sources~\cite{MUSIC_LWA_non_correlated, gazzah}. However, they fail to handle coherent sources because the rank of the source covariance matrix collapses, and the LWA steering matrix does not exhibit a Vandermonde structure~\cite{rida}. This prevents the use of traditional techniques such as spatial smoothing (SSP) to restore the rank. More recently, the Spatially Filtered Interpolation (SFI) method has been introduced~\cite{rida}, which sectorizes the FoV according to the LWA frequency beam-scanning capability and applies the spatial filtering (SF) method to suppress out-of-sector interference before interpolation. While effective, SFI still suffers from numerical instability. These limitations have motivated alternative strategies, and recent work by Zhu et al.~\cite{zhu2025sbllwa} has highlighted the potential of Bayesian learning approaches for LWA-based DoA estimation.

Sparse Bayesian Learning (SBL) has emerged as a robust alternative for DoA estimation, since it does not rely on a Vandermonde model and can resolve multiple closely spaced or coherent sources even with limited snapshots~\cite{chapter11,tipping2001sparse}. For LWAs, SBL directly processes the frequency-dependent steering response~\cite{zhu2025sbllwa}. Two main variants are typically used, on-grid SBL, which estimates source activity on a fixed angular grid, and off-grid SBL, which improves accuracy by modeling small angular deviations through a first-order Taylor expansion of the steering vector~\cite{yang2013offgrid,dai2017offgrid}. In contrast, gridless approaches such as atomic norm minimization remain inapplicable to LWAs because they require a Vandermonde structure~\cite{chapter11}.

Building on these advances, this work proposes a sector-wise SF-based method that integrates on-grid and off-grid SBL for DoA estimation with single-beam LWAs. The FoV is divided into angular sectors based on the frequency–angle mapping, and only the frequency components whose main beams fall within a sector are selected, which reduces basis mismatch and computational load. Subsequently, in each sector, on-grid SBL estimates the source directions, while an off-grid refinement step compensates for residual grid mismatch to enhance resolution. The method combines the robustness of SBL with the hardware efficiency of LWAs, enabling accurate coherent-source estimation without interpolation or subspace decomposition.

The paper is organized as follows. Section~\ref{sec:antenna_model} introduces the LWA model, while Section~\ref{sec:system_model} presents the system model. The proposed SF-based method integrating on-grid and off-grid SBL is described in Section~\ref{sec:sfi_sbl}. Simulation results are reported in Section~\ref{sec:results}, and finally, conclusions are drawn in Section~\ref{sec:conclusion}.
\section{Leaky-Wave Antenna Model}
\label{sec:antenna_model}

In this work, we consider a 1D periodic unidirectional leaky-wave waveguide of physical length \( l_a \) and spatial period \( p \). The LWA supports a complex propagation mode that radiates energy into free space, producing a far-field pattern that varies with frequency. The far-field response of the LWA at frequency \( f \) and observation angle \( \theta \) is modeled as~\cite{LWAbook}:
\begin{equation}
    a_f(\theta) = l_a \, e^{-j (k_z - k_0 \sin\theta) \frac{l_a}{2}} \text{sinc}\left[(k_z - k_0 \sin\theta) \frac{l_a}{2}\right],
    \label{eq:lwa_steering}
\end{equation}
where \( k_0 = \frac{2\pi f}{c} \) is the free-space wavenumber, \( c \) is the speed of light, and \( k_z = \beta - j\alpha \) is the complex wavenumber along the waveguide. Here, \( \alpha \) accounts for leakage loss (i.e., radiation), and \( \beta \) represents the guided-mode phase constant.
Assuming the modulation of the waveguide generates a single fast spatial harmonic (typically the \( -1 \)st), the guided-mode phase constant can be expressed as:
\begin{equation}
    \beta = \beta_0 - \frac{2\pi}{p},
\end{equation}
where \( \beta_0 \) is the unperturbed phase constant of the dielectric-filled waveguide. Under fundamental mode operation, \( \beta_0 \) can be approximated by:
\begin{equation}
    \beta_0 = k_0 \sqrt{\epsilon_r} \sqrt{1 - \left({f_c}/{f}\right)^2},
    \label{eq:beta0}
\end{equation}
with \( \epsilon_r \) denoting the relative permittivity of the dielectric medium and \( f_c = \frac{c}{2 W_g \sqrt{\epsilon_r}} \) representing the waveguide’s cutoff frequency, determined by its width \( W_g \).

For radiation to occur, the condition \( -k_0 \leq \beta \leq k_0 \) must hold. The main beam direction as a function of frequency is then given by~\cite{LWAbook}:
\begin{equation}
    \theta_0(f) = \sin^{-1}\left(\frac{\beta}{k_0}\right),
    \label{eq:beam_angle}
\end{equation}
showing how the radiation angle \( \theta_0 \) varies with frequency. This frequency-to-angle mapping enables beam steering over a wide field of view using a single physical antenna element.

\begin{figure}[ht]
    \centering
        \includegraphics[width=\linewidth]{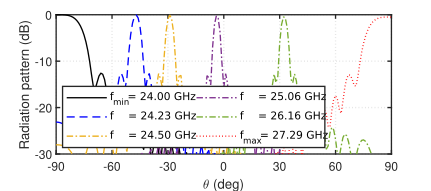}
        \caption{Normalized radiation pattern of the LWA.}
        \label{pattern}

\end{figure}
In this study, the LWA is designed with a relative permittivity \( \epsilon_r = 10.2 \), waveguide width \( W_g = 2.1\,\text{mm} \), modulation period \( p = 5.5\,\text{mm} \), total length \( l_a = 20\,\text{cm} \), and leakage factor \( \alpha/k_0 = 0.01 \). These parameters enable full angular scanning from \( -90^\circ \) to \( +90^\circ \) as the frequency sweeps from \( f_{\min} = 24 \,\text{GHz} \) to \( f_{\max} = 27.29 \,\text{GHz} \). The 3-dB beamwidth is approximately \(10^\circ \) near broadside and gradually broadens toward end-fire. This frequency-dependent beam-steering property forms the foundation of the SF-based method described in the following sections.

\section{Signal Model and Problem Formulation}
\label{sec:system_model}

Consider \( K \) narrowband plane wave signals impinging on the LWA from distinct directions \( \theta_k \), where \( k = 1, \dots, K \). Each source transmits a multi-carrier signal (e.g., OFDM) spanning \( N \) uniformly spaced frequency points between \( f_{\min} \) and \( f_{\max} \). We assume the source amplitude is constant across subcarriers, i.e., \( s_{k_n} = s_k \) for all \( n \) and fixed \( k \). The frequency vector is defined as:
\begin{equation}
    \mathbf{f} = \left[ f_{\min}, f_{\min} + \Delta f, \dots, f_{\max} \right]^T,
    \label{eq:freq_vector}
\end{equation}
where $\Delta f = \frac{f_{\text{max}} - f_{\text{min}}}{N - 1}$ and $f_n$ is the $n^{th}$ element of the frequency vector $\mathbf{f}$. The following frequency-domain system model can be used to express the received signal~\cite{MUSIC_MULTI}:
\begin{equation}
    \mathbf{y}[t] = \mathbf{A}\mathbf{x}[t] + \mathbf{n}[t],
    \label{eq:recieved}
\end{equation}
where $\mathbf{y}[t]\in\mathbb{C}^{N\times 1}$ is the received data vector,
with $t=1,2,\dots,T$ denoting the $t$-th snapshot among $T$.
The steering matrix is
$\mathbf{A}=[\mathbf{a}(\theta_1),\dots,\mathbf{a}(\theta_K)]\in\mathbb{C}^{N\times K}$,
with steering vectors
\(
\mathbf{a}(\theta) = [a_{f_1}(\theta),\,a_{f_2}(\theta),\,\dots,\,a_{f_N}(\theta)]^{\top},
\)
where $a_{f_n}(\theta)$ is the LWA response at frequency $f_n$ given in~\eqref{eq:lwa_steering}.
The source vector is $\mathbf{x}[t]\in\mathbb{C}^{K\times 1}$,
and the additive white Gaussian noise (AWGN) vector is
$\mathbf{n}[t]\in\mathbb{C}^{N\times 1}$ with independent entries,
zero mean, and covariance $\sigma^2\mathbf{I}_N$.
The columns of $\mathbf{A}$ represent the LWA responses to incoming
plane waves at directions $\theta_i$, $i=1,\dots,K$.
We assume $N>K$.

% with the received data vector $\mathbf{y} \in \mathbb{C}^{N \times 1}$, the LWA steering matrix is defined as $\mathbf{A} = [\mathbf{a}(\theta_1), \mathbf{a}(\theta_2), \dots, \mathbf{a}(\theta_K)]$, where $\mathbf{a}(\theta) = [a_1(\theta), a_2(\theta), \dots, a_N(\theta)]^T$. The source vector is $\mathbf{s} \in \mathbb{C}^{K \times 1}$, and the additive white Gaussian noise (AWGN) vector is $\mathbf{n} \in \mathbb{C}^{N \times 1}$ with independent components, zero mean, and covariance matrix $\sigma^2 \mathbf{I}_N$. The columns of $\mathbf{A}$ represent the LWA responses to incoming plane waves with directions of arrival $\theta_i,\, i=1,\dots,K$, as described in~\eqref{eq:lwa_steering}. In addition, it is assumed that $N > K$.

The covariance matrix of the received signal vector (\ref{eq:recieved}) can be written as:
\begin{equation}
     \mathbf{R} = E[ \mathbf{y} \mathbf{y}^H]
     = \mathbf{A}\mathbf{R}_x \mathbf{A}^H + \sigma^2 \mathbf{I}_N ,
     \label{ref:lwa_covariance}
\end{equation}

For coherent sources, subspace-based methods fail due to rank deficiency and the non-Vandermonde structure of the LWA response~\cite{rida}. SBL methods can overcome these issues by directly processing the LWA steering matrix~\cite{chapter11}, except gridless SBL methods, which still require a Vandermonde structure. Consequently, we propose a sector-wise SF-based method combined with SBL to enable robust DoA estimation.

\section{SBL with spatially-filtered}
\label{sec:sfi_sbl}
Following~\cite{rida}, we propose an SF-SBL framework that exploits the LWA frequency–angle mapping to divide the FoV into $L$ angular sectors indexed by $\ell = 1,\dots,L$. For each sector, a set of $P$ candidate DoAs is defined as\(\
\Theta^{(\ell)} = \big[ \theta^{(\ell)}_1, \, \theta^{(\ell)}_1 + \Delta\theta^{(\ell)}, \, \dots, \, \theta^{(\ell)}_P \big],
\) where $\Delta\theta^{(\ell)}$ is the angular step in sector $\ell$.  
To spatially filter out sources outside sector~$\ell$, a frequency selection is applied by exploiting the LWA frequency--angle mapping. Specifically, a subset of $N_s$ frequency samples is extracted from the received vector $\mathbf{y}$ as
\begin{equation}
\mathbf{y}_s^{(\ell)} =
\begin{bmatrix}
y_{i^{(\ell)}} , \dots, y_{i^{(\ell)}+N_s-1}
\end{bmatrix}^T ,
\label{eq:filtered_y}
\end{equation}
where \( i^{(\ell)} \) is the \( i^{(\ell)} \)-th frequency component \( f_{i^{(\ell)}} \) that corresponds to the LWA beam direction \( \theta^{(\ell)}_1 \) within sector \( {(\ell)} \). And, the frequency component \( f_{\,i^{(\ell)} + N_s - 1} \) corresponds to the LWA beam direction \( \theta^{(\ell)}_P \), based on the relation given in (\ref{eq:beam_angle}). 

\subsection{Sector width selection}

The 3-dB beamwidth of the LWA at frequency $f$ can be approximated by~\cite{balanis2016, jackson2012}
\begin{equation}
    B_\theta(f) \approx \frac{180}{\pi} \, \frac{\lambda(f)}{l_a \cos \theta_0(f)},
    \label{eq:beamwidth}
\end{equation}
where $\lambda(f) = c/f$ is the wavelength, $l_a$ is the aperture length, and $\theta_0(f)$ is the main beam direction. The maximum beamwidth across the band is
\begin{equation}
    B_{\theta,\max} = \max_f B_\theta(f).
    \label{eq:max_beamwidth}
\end{equation}

To ensure that each sector fully contains the main lobe at all relevant frequencies, the sector width $w_\theta$ is chosen such that
\begin{equation}
    w_\theta \geq B_{\theta,\max}.
    \label{eq:sector_width}
\end{equation}

This condition ensures that the main-lobe energy of each frequency is fully contained within a single sector, avoiding leakage across adjacent sectors. In practice, we use
\begin{equation}
    w_\theta = \gamma \, B_{\theta,\max},
    \label{eq:sector_width_gamma}
\end{equation}
with $\gamma > 1$ a design parameter trading robustness against complexity.

\subsection{SF-based On-Grid SBL}

After applying spatial filtering, we adopt the SBL framework. In this section, we apply the on-grid SBL method to each sector, where the data model is given by:
\begin{equation}
    \mathbf{Y}_s^{(\ell)} = \mathbf{A}_s^{(\ell)} \mathbf{X}^{(\ell)} + \mathbf{N}^{(\ell)},
\end{equation}
where $\mathbf{Y}_s^{(\ell)} \in \mathbb{C}^{N_s \times T}$ contains 
the $N_s$ frequency samples of sector~$\ell$ across $T$ snapshots, 
$\mathbf{A}_s^{(\ell)} \in \mathbb{C}^{N_s \times P}$ is the steering matrix, 
$\mathbf{X}^{(\ell)} \in \mathbb{C}^{P \times T}$ is the row-sparse source matrix, 
and $\mathbf{N}^{(\ell)}$ is noise.  
Each row $\mathbf{x}_{p,:}^{(\ell)} \sim \mathcal{CN}(\mathbf{0}, \gamma_p \mathbf{I}_T)$, 
with $\gamma_p \geq 0$, while the noise variance $\sigma^2$ is assigned a weak Gamma prior.

The EM updates alternate between posterior inference and hyperparameter refinement.  
Following~\cite{zhu2025sbllwa}, the sparsity update is
\begin{equation}
    \gamma_p \;\leftarrow\; 
    \frac{1}{2\varsigma}\!\left(
    \sqrt{T^2 + 4\varsigma(\|\mathbf{u}_{p,:}\|_2^2 + T\,\boldsymbol{\Sigma}_{pp})} - T
    \right),
    \label{eq:gamma_update}
\end{equation}
with $\varsigma=10^{-2}$, $\mathbf{u}_{p,:}$ the $p$th row of
$\mathbf{U}=\sigma^{-2}\boldsymbol{\Sigma}\mathbf{A}_s^{(\ell)H}\mathbf{Y}_s^{(\ell)}$, 
and $\boldsymbol{\Sigma}_{pp}$ the $p$th diagonal element of
\[
\boldsymbol{\Sigma} = \boldsymbol{\Gamma} - 
\boldsymbol{\Gamma}\mathbf{A}_s^{(\ell)H}\mathbf{C}^{-1}\mathbf{A}_s^{(\ell)}\boldsymbol{\Gamma},
\quad
\mathbf{C} = \sigma^2 \mathbf{I}_{N_s} + \mathbf{A}_s^{(\ell)}\boldsymbol{\Gamma}\mathbf{A}_s^{(\ell)H}.
\]

The noise variance is updated as
\begin{equation}
    \sigma^2 \;\leftarrow\;
    \frac{\|\mathbf{Y}_s^{(\ell)} - \mathbf{A}_s^{(\ell)} \mathbf{U}\|_F^2 
    + \sigma^2 T \sum_{p=1}^P (1 - \gamma_p^{-1}\boldsymbol{\Sigma}_{pp}) + d}
    {N_s T + c - 1},
    \label{eq:sigma_update}
\end{equation}
with $(c,d)=(10^{-4},10^{-4})$. Iterating \eqref{eq:gamma_update}–\eqref{eq:sigma_update} 
drives inactive rows of $\mathbf{X}^{(\ell)}$ to zero, revealing the DoAs.  

% ------------------------------------------------------

\subsection{SF-based Off-Grid SBL}
While the on-grid formulation assumes that the true DoAs coincide with the predefined grid $\{\vartheta_p\}_{p=1}^P$, in practice sources rarely align perfectly, which leads to basis mismatch. To mitigate this, we adopt the off-grid SBL framework~\cite{yang2013offgrid}, in which each direction is parameterized as\begin{equation}
    \theta_p = \vartheta_p + \beta_p,
\end{equation}
with small offset $\beta_p$. A first-order Taylor expansion of the steering vector yields
\begin{equation}
    \mathbf{Y}_s^{(\ell)} \;\approx\;
    \big(\mathbf{A}_s^{(\ell)} + \mathbf{B}_s^{(\ell)}\operatorname{diag}(\boldsymbol{\beta})\big)
    \mathbf{X}^{(\ell)} + \mathbf{N}^{(\ell)},
\end{equation}
where $\mathbf{B}_s^{(\ell)}$ contains angular derivatives and 
$\boldsymbol{\beta}=[\beta_1,\dots,\beta_P]^T$.  

The EM updates for $\{\gamma_p\}$ and $\sigma^2$ remain as in the on-grid case.  
The off-grid offsets $\{\beta_p\}$ are then refined following~\cite{yang2013offgrid} as
\begin{equation}
    \beta_p \;\leftarrow\; 
    \frac{\Re\!\{\mathbf{a}_s^{(\ell)}(\vartheta_p)^H \boldsymbol{\Sigma}\,\mathbf{b}_s^{(\ell)}(\vartheta_p)\}}
         {\Re\!\{\mathbf{b}_s^{(\ell)}(\vartheta_p)^H \boldsymbol{\Sigma}\,\mathbf{b}_s^{(\ell)}(\vartheta_p)\}},
    \qquad p=1,\dots,P,
\end{equation}
where $\mathbf{a}_s^{(\ell)}(\vartheta_p)$ and $\mathbf{b}_s^{(\ell)}(\vartheta_p)$ 
are the steering vector and its angular derivative at $\vartheta_p$, 
and $\boldsymbol{\Sigma}$ is the posterior covariance of the coefficients.

By alternating between hyperparameter updates and $\beta$ refinement, the algorithm 
converges to sparse estimates $\hat{\theta}_p = \vartheta_p + \hat{\beta}_p$ for all 
active indices. The complete procedure of the SF-OnGrid-SBL and SF-OffGrid-SBL methods for single beam LWA is outlined in Algorithm~\ref{alg:sfi_sbl_sector_algx}.

\begin{algorithm}[t]
\caption{Spatially filtered on-/off-grid SBL applied within sector $\ell$}
\label{alg:sfi_sbl_sector_algx}
\begin{algorithmic}[1]
\Require Sector $\ell$; $\mathbf{Y}_s^{(\ell)} \in \mathbb{C}^{N_s\times T}$; 
on-grid steering matrix $\mathbf{A}_s^{(\ell)}$; (off-grid) derivative matrix $\mathbf{B}_s^{(\ell)}$; grid $\{\vartheta_p\}$.
\State Initialize $\gamma_p \!\leftarrow\! \gamma_0>0$, $\beta_p \!\leftarrow\! 0$; noise precision prior $\sigma^{-2} \sim \mathrm{Gamma}(c,d)$ with $c=d=10^{-4}$~\cite{tipping2001sparse}.
\Repeat
  \State \textbf{E-step:} form effective dictionary $\mathbf{\Phi}(\boldsymbol{\beta}) \leftarrow \mathbf{A}_s^{(\ell)} + \mathbf{B}_s^{(\ell)} \operatorname{diag}(\boldsymbol{\beta})$.
  \State Update posterior covariance: $\boldsymbol{\Sigma} \leftarrow \big(\operatorname{diag}(\boldsymbol{\gamma})^{-1} + \tfrac{1}{\sigma^2}\mathbf{\Phi}^H\mathbf{\Phi}\big)^{-1}$.
  \For{$t=1,\dots,T$}
     \State Compute posterior mean: $\boldsymbol{\mu}_t \leftarrow \tfrac{1}{\sigma^2}\boldsymbol{\Sigma}\mathbf{\Phi}^{H}\mathbf{y}^{(\ell)}_{s}[t]$.
  \EndFor
  \State \textbf{M-step (sparsity):} update $\gamma_p \leftarrow \tfrac{1}{T}\sum_{t=1}^T |\mu_{p,t}|^2 + \Sigma_{pp}$~\cite{chapter11}.
  \State \textbf{M-step (noise):} update $\sigma^2$ by evidence maximization with $(c,d)=(10^{-4},10^{-4})$~\cite{tipping2001sparse}.
  \If{off-grid enabled}
     \State \textbf{Off-grid refinement:} update each $\beta_p$ using the OGSBI linearized step~\cite{yang2013offgrid}.
  \EndIf
  \State Prune small $\gamma_p$; optionally merge adjacent active components.
\Until{convergence}
\State \textbf{Return:} estimated DoAs $\widehat{\theta}_p = \vartheta_p + \widehat{\beta}_p$ for active indices $p$; deduplicate across overlapping sectors.
\end{algorithmic}
\end{algorithm}

\section{Simulation Results}
\label{sec:results}
This section presents the simulation results of the proposed methods. 
All experiments were carried out on a standard laptop (Intel Core i5, MATLAB implementation). 
The single-beam LWA parameters are given in Section~\ref{sec:antenna_model}, with the number 
of frequency samples fixed to $N=100$. Unless otherwise stated, the field of view was set to 
$\Theta_{\text{FoV}}=\{-90^\circ,-90^\circ+\Delta\theta,\ldots,90^\circ\}$ with a grid step 
of $\Delta\theta=1^\circ$, used both for coarse search and refinement. Each run used $T=100$ 
snapshots, and performance metrics were averaged over $I=100$ Monte Carlo trials.  

The angular domain was divided into $L=6$ sectors, each spanning $w_\theta=30^\circ$. 
This choice ensures compatibility with the maximum 3-dB beamwidth of the antenna, 
$B_{\theta,\max}=10.66^\circ$, and corresponds to $\gamma \approx 2.8$. 
The frequency samples in each sector were assigned according to the frequency–angle mapping 
inherent to the LWA.  

For benchmarking, four algorithms were considered. The proposed approaches are 
SF-OnGrid-SBL and SF-OffGrid-SBL. As references, the Cropped-SVD SBL method~\cite{zhu2025sbllwa} 
was applied over the entire FoV, and the SFI+SSP+MUSIC framework~\cite{rida} was implemented 
with the same $L=6$ sectors and $30^\circ$ sector width. In the SFI+SSP+MUSIC setup, the virtual 
array was defined with $V=N_s=16$ elements, $P=14$ virtual angles, and an SSP subarray length 
of $F=0.6V$.
\begin{figure}[t]
    \centering
    \includegraphics[width=\linewidth]{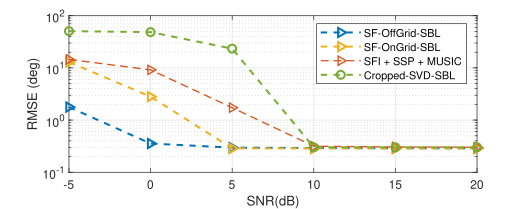}
    \caption{RMSE performance versus SNR for three coherent sources at DoAs $[10.3^\circ,\,15.7^\circ,\,20.7^\circ]$.}
    \label{fig:rmse_analysis}
\end{figure}

\begin{figure}[t]
    \centering
    \includegraphics[width=\linewidth]{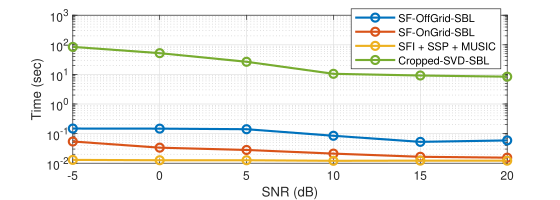}
    \caption{Computational time of the compared algorithms versus SNR for three coherent sources at DoAs $[10.3^\circ,\,15.7^\circ,\,20.7^\circ]$.}
    \label{fig:time_analysis}
\end{figure}

% \begin{figure}
%     \centering
%     \begin{subfigure}{\linewidth}
%         \centering
%         \includegraphics[width=\linewidth]{RMSE_30_sector_ICASSP.eps}
%         \caption{RMSE performance versus SNR.}
%         \label{fig:rmse_analysis}
%     \end{subfigure}

%     \vspace{4pt}

%     \begin{subfigure}{\linewidth}
%         \centering
%         \includegraphics[width=\linewidth]{time_run.eps}
%         \caption{Computational time versus SNR.}
%         \label{fig:time_analysis}
%     \end{subfigure}
%     \caption{Performance of the compared algorithms for three coherent sources at DoAs $[10.3^\circ,\,15.7^\circ,\,20.7^\circ]$.}
%     \label{fig:results}
% \end{figure}
Fig.~\ref{fig:rmse_analysis} shows the RMSE performance for three coherent sources at DoAs $[10.3^\circ,\,15.7^\circ,\,20.7^\circ]$ across different SNR levels. The proposed SF-OffGrid-SBL consistently achieves the lowest RMSE, showing strong robustness against grid mismatch. The SF-OnGrid-SBL improves as the SNR increases, but its performance remains limited at low SNR. The classical SFI+SSP+MUSIC method provides moderate accuracy, with performance stabilizing at higher SNR values. Finally, the Cropped-SVD SBL baseline yields the worst accuracy among all methods, although it performs reasonably well at high SNR. These results confirm the effectiveness and robustness of the proposed approach across a SNR range.

Fig.~\ref{fig:time_analysis} reports the computational time versus SNR. The Cropped-SVD SBL method requires significantly higher runtime, mainly due to the full FoV processing and large dictionary size. In contrast, both SF-OnGrid-SBL and SF-OffGrid-SBL methods considerably reduce the execution time by exploiting sector-wise processing. The SFI+SSP+MUSIC method remains the most computationally efficient, although at the cost of reduced estimation accuracy. Overall, the results highlight a favorable trade-off achieved by the proposed SF-OffGrid-SBL, combining high accuracy with competitive computational efficiency.
\section{Conclusion}
\label{sec:conclusion}
This paper presents a SF-SBL framework for DoA estimation with single-beam LWAs.
By exploiting the frequency–angle mapping, the SF-OnGrid-SBL and SF-OffGrid-SBL reduce complexity and avoid numerical interpolation instabilities while handling coherent sources.
Simulations show that SF-OffGrid-SBL achieves the lowest RMSE with competitive runtime, outperforming existing methods.
Future work will extend the approach to multi-beam LWAs and validate it on measured antenna data.

% \bibliographystyle{IEEEtran}
% \bibliography{refs}

\section*{Acknowledgment}
This work was supported by the ANR BeSensiCom project (Grant ANR-22-CE25-0002) of the French Agence Nationale de la Recherche, 
the National Key Research and Development Program of China under Grant~2024YFE0105400, 
and carried out in the framework of COST Action CA20120 INTERACT.

\end{document}